\def\abbrename{{\bfseries Abbreviations \runinend}}

\documentclass[global,referee]{svjour}
\usepackage{graphicx}

\begin{document}
\titlerunning{Scale-dependent dynamic SGS models for LES of NBL}
\authorrunning{Anderson et al.} 
\title{Comparison of Two Scale-Dependent Dynamic Subgrid-Scale Models 
for Simulations of Neutrally Buoyant Shear-Driven Atmospheric Boundary
Layer Flows}
\author{W.~C. Anderson\inst{1} \and S.~Basu
\inst{2}\thanks{\emph{Corresponding address:} S.~Basu, 
Atmospheric  Science  Group,  Department  of Geosciences,  Texas  Tech
University, Lubbock, TX 79409, USA, \email{sukanta.basu@ttu.edu}}
\and C.~W.~Letchford \inst{1}
} 
%
%
\institute{Wind Science and Engineering Research Center, Texas Tech University,
Lubbock, TX  79409, USA \and Atmospheric Science  Group, Department of
Geosciences, Texas Tech University, Lubbock, TX 79409, USA}
%

%
\maketitle

\begin{abstract}
A  new  scale-dependent dynamic  subgrid-scale  (SGS)  model based  on
Kolmogorov's  scaling  hypothesis  is  presented. This  SGS  model  is
utilized  in  large-eddy simulation  of  a  well-known  case study  on
shear-driven neutral atmospheric boundary layer flows. The results are
compared comprehensively with an alternate scale-dependent dynamic SGS
model  based on  the popular  Smagorinsky closure.   Our  results show
that, in  the context of this particular  problem, the scale-dependent
dynamic modeling approach is  extremely useful, and reproduces several
establised results  (e.g., the  surface layer similarity  theory) with
fidelity. Results from both the SGS base models are generally in close
agreement,  although   we  find   a  consistent  superiority   of  the
Smagorinsky-based SGS model for predicting the inertial range scaling 
of spectra.
\end{abstract}
 

\abbrename{ABL - Atmospheric boundary layer; LES - Large-eddy simulation; 
SGS -  Subgrid scale; NBL -  Neutral boundary layer;  TKE - Turbulence
kinetic energy.}

\section{Introduction}\label{Sec1}

The dynamic subgrid-scale (SGS) modeling approach of Germano et al.
\cite{germ91}  has  been quite  successful  in large-eddy  simulations
(LESs) of various engineering  flows \cite{pope04}.  In this approach,
one dynamically computes the values of the unknown SGS coefficients at
every time and position in the flow. By looking at the dynamics of the
flow at  two different resolved scales, and  assuming scale similarity
as well as scale invariance  of the SGS coefficients, one can optimize
their values. Thus, the dynamic  modeling approach avoids the need for
{\it a  priori} specification and  tuning of the SGS  coefficients.  A
recent  study  \cite{meye05}  based  on  extensive  database  analysis
further suggests that the dynamic modeling approach closely reproduces
the minimal  simulation error  strategy (termed as  optimal refinement
strategy), which is highly desirable in turbulence modeling.

In  atmospheric  boundary  layer  (ABL) turbulence,  where  shear  and
stratification   and  associated   flow   anisotropies  are   (almost)
ubiquitous, the  inherent scale-invariance assumption  of the original
dynamic modeling approach breaks down.  Port\'{e}-Agel et al.
\cite{port00} relaxed this assumption and introduced a scale-dependent
dynamic modeling approach in which the SGS coefficients are assumed to
vary  as powers  of the  LES filter  width ($\Delta_f$).   The unknown
power-law  exponents, and  subsequently the  SGS coefficients,  can be
determined in a self-consistent manner by filtering at three levels
\cite{port00,port04}.  In  the simulations of  neutral boundary layers
(NBLs),  the scale-dependent dynamic  SGS model  was found  to exhibit
appropriate  dissipation   behavior  and  more   accurate  spectra  in
comparison to the original (scale-invariant) dynamic model
\cite{port00,port04}.   Recently the scale-dependent  dynamic modeling
approach  was  modified  and  extended by  incorporating  a  localized
averaging  technique   in  order  to   simulate  intermittent,  patchy
turbulence in the stably stratified flows
\cite{basu06a,basu06b}.   In  parallel,  scale-dependent  dynamic  SGS
models based on  Lagrangian averaging over fluid flow  path lines were
developed   by  Bou-zeid   et   al.   \cite{bouz05}   and  Stoll   and
Port\'{e}-Agel  \cite{stol06} to  simulate neutrally  stratified flows
over heterogeneous surfaces.

The scale-dependent dynamic modeling  approach and its variants so far
always used the popular eddy-viscosity formulation of Smagorinsky
\cite{smag63} as the SGS base model.  However, this SGS model assumes
that  the energy  dissipation rate  equals the  SGS  energy production
rate.  In order to avoid this strong assumption, Wong and Lilly
\cite{wong94} proposed  a new SGS model based  on Kolmogorov's scaling
hypothesis.  A  dynamic version  of the Wong-Lilly  SGS model  to some
extent outperformed  the dynamic  Smagorinsky model in  simulations of
the  buoyancy-driven   Rayleigh-B\'{e}nard  convection  \cite{wong94}.
Furthermore,  the  dynamic  Wong-Lilly  SGS model  is  computationally
inexpensive in  comparison to the dynamic Smagorinsky  SGS model.  The
combination  of  lesser  assumptions  and cheaper  computational  cost
certainly make the  Wong-Lilly model an attractive SGS  base model for
LES.  Therefore  it is  of interest to  explore if the  Wong-Lilly SGS
model or its variants are capable of simulating different flow regimes
of  the ABL.   It   is  generally  agreed  upon  that   in  comparison  to
buoyancy-driven flows, large-eddy simulations of shear-driven boundary
layer flows are  far more challenging. Thus, in  the present study, we
focus  on  neutrally  buoyant  shear-driven  ABL flow.   In  order  to
realistically  account  for  the  near-wall shear  effects,  we  first
formulate a  locally averaged  scale-dependent dynamic version  of the
Wong-Lilly SGS model (henceforth  LASDD-WL, see Appendix for details).
Then,  we comprehensively  compare  its performance  with the  locally
averaged scale-dependent dynamic  Smagorinsky (hereafter LASDD-SM) SGS
model earlier developed by Basu and Port\'{e}-Agel
\cite{basu06a}.

The structure  of this paper is  as follows. In Section  2, we briefly
provide the  technical details of a case  study. Extensive comparisons
(in terms of the  similarity theory, spectra, and flow visualizations)
between the LASDD-WL and LASDD-SM  SGS models are performed in Section
3. Finally, concluding remarks are made in Section 4.

\section{Description of Simulations}\label{Sec2}

In this work,  we perform large-eddy simulations of  a turbulent Ekman
layer (i.e.,  pure shear flow with a  neutrally stratified environment
in  a rotating system)  utilizing the  LASDD-SM \cite{basu06a,basu06b}
and LASDD-WL  (see Appendix) SGS  models.  Both these  simulations are
identical  in terms  of  initial conditions,  forcings, and  numerical
specifications  (e.g.,  time  integration, grid  spacing).   Technical
details of  our LES code and  the LASDD-SM SGS  modeling approach have
been described  in detail in  \cite{basu06a} and will not  be repeated
here for brevity.

The selected case study is  similar to that of the LES intercomparison
study  by Andr\'{e}n  et al.   \cite{andr94}.  The  simulated boundary
layer    is   driven    by    an   imposed    geostrophic   wind    of
$\left(U_g,V_g\right)  = \left(10,0\right)$  ms$^{-1}$.   The Coriolis
parameter  is equal  to  $f_c =  10^{-4}$  s$^{-1}$, corresponding  to
latitude $45^\circ$ N.  The computational domain size is: $L_x = L_y =
4000$ m and  $L_z = 1500$ m.  This domain is  divided into $N_x \times
N_y  \times  N_z = 40 \times 40 \times 40$ nodes  (i.e.,  $\Delta_x =  \Delta_y  =  100$ m,  and
$\Delta_z  = 38.5$  m). The  motivation behind  the selection  of this
coarse grid-resolution is two-fold.  Primarily it allows us to perform
a direct  comparison with the  results from \cite{andr94},  which used
almost   the   same   grid-resolutions.   More   importantly,   coarse
grid-resolution enables us to identify the strengths and/or weaknesses
of different  SGS models, as well  as, to underscore  their impacts on
large-eddy simulations.  The simulations are run for a period of $10
\times f_c^{-1}$ (i.e.,  100,000 s), with time steps of  2 s. The last
$3 \times f_c^{-1}$ interval is used to compute statistics.  A passive
scalar  is  introduced  in  the  flows by  imposing  a  constant  flux
($\overline{wc}_0$) of $10^{-3}$ kg  m$^{-2}$ s$^{-1}$ at the surface.
The lower boundary condition  is based on the Monin-Obukhov similarity
theory with a surface roughness length of $z_\circ = 0.1$ m.

\section{Results and Discussions}\label{Sec3}

In this  section, we report the  results of the  LASDD-SM and LASDD-WL
SGS models-based  simulations and compare  them with results  from the
intercomparison   study   \cite{andr94},   wherever  possible.    This
particular case  (without the inclusion  of passive scalars)  was also
simulated by  Kosovi\'{c} \cite{koso97}  using a nonlinear  SGS model,
and  recently   by  Chow  et   al.   \cite{chow05},  who   utilized  a
sophisticated hybrid  SGS model.  Our  simulations show that  both the
LASDD SGS models perform very  well, and the results are comparable to
the past studies.

Temporal evolution  of the surface  friction velocity ($u_*$)  is very
similar in  both the  simulations (not shown).   The average  value of
$u_*$ during  the last $3  \times f_c^{-1}$ interval  is approximately
0.44 ms$^{-1}$ in  the case of the LASDD-SM  model. The LASDD-WL model
produces   a   marginally  higher   value   (0.454  ms$^{-1}$).    The
corresponding  values  found  in  \cite{andr94} are:  0.425  ms$^{-1}$
(Moeng), 0.448 ms$^{-1}$ (Mason - backscatter), 0.402 ms$^{-1}$ (Mason
- non-backscatter), 0.402 ms$^{-1}$  (Nieuwstadt), and 0.425 ms$^{-1}$
(Schumann).

\begin{figure}[ht]
\centerline{\includegraphics[width=2.25in]{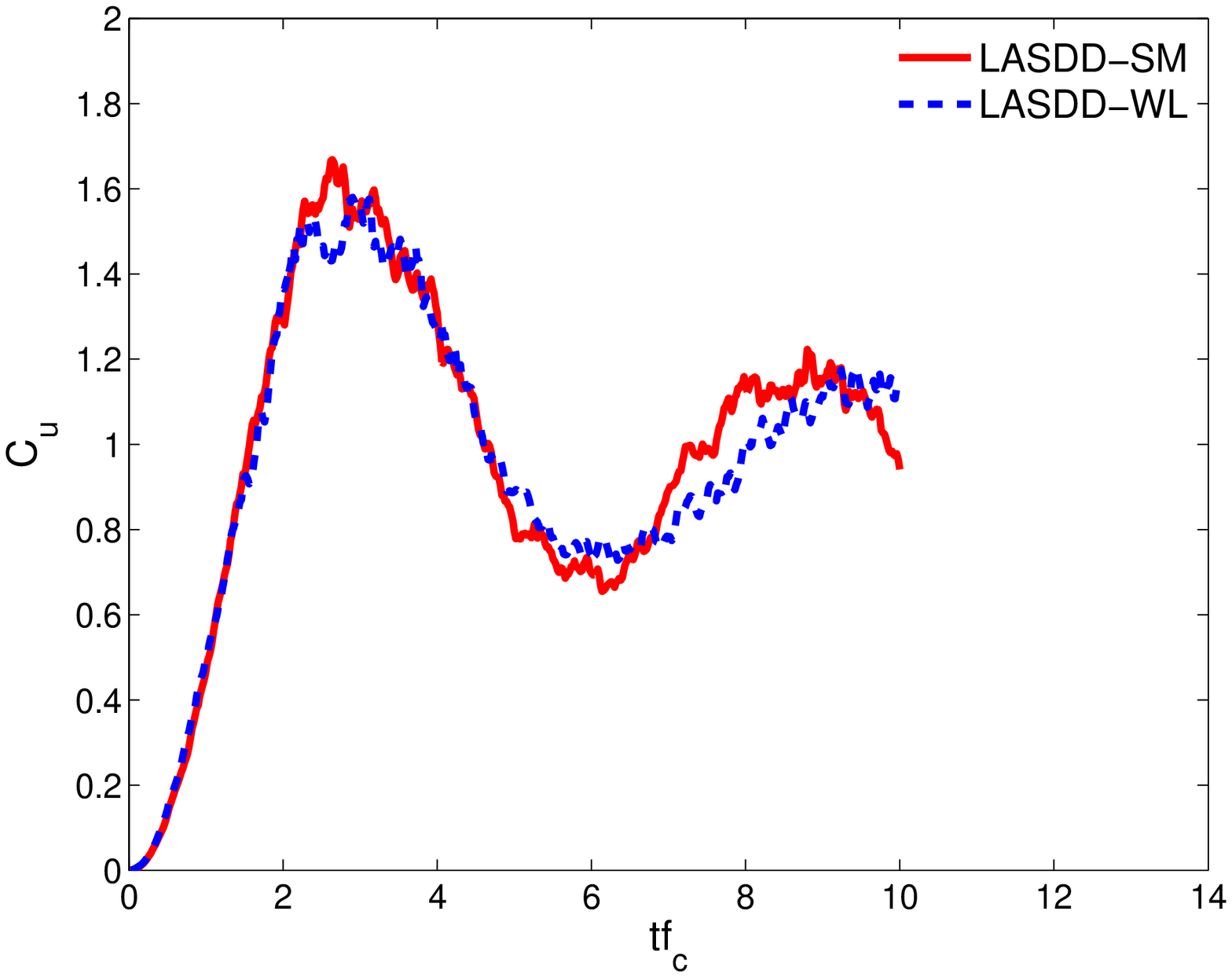}
\includegraphics[width=2.25in]{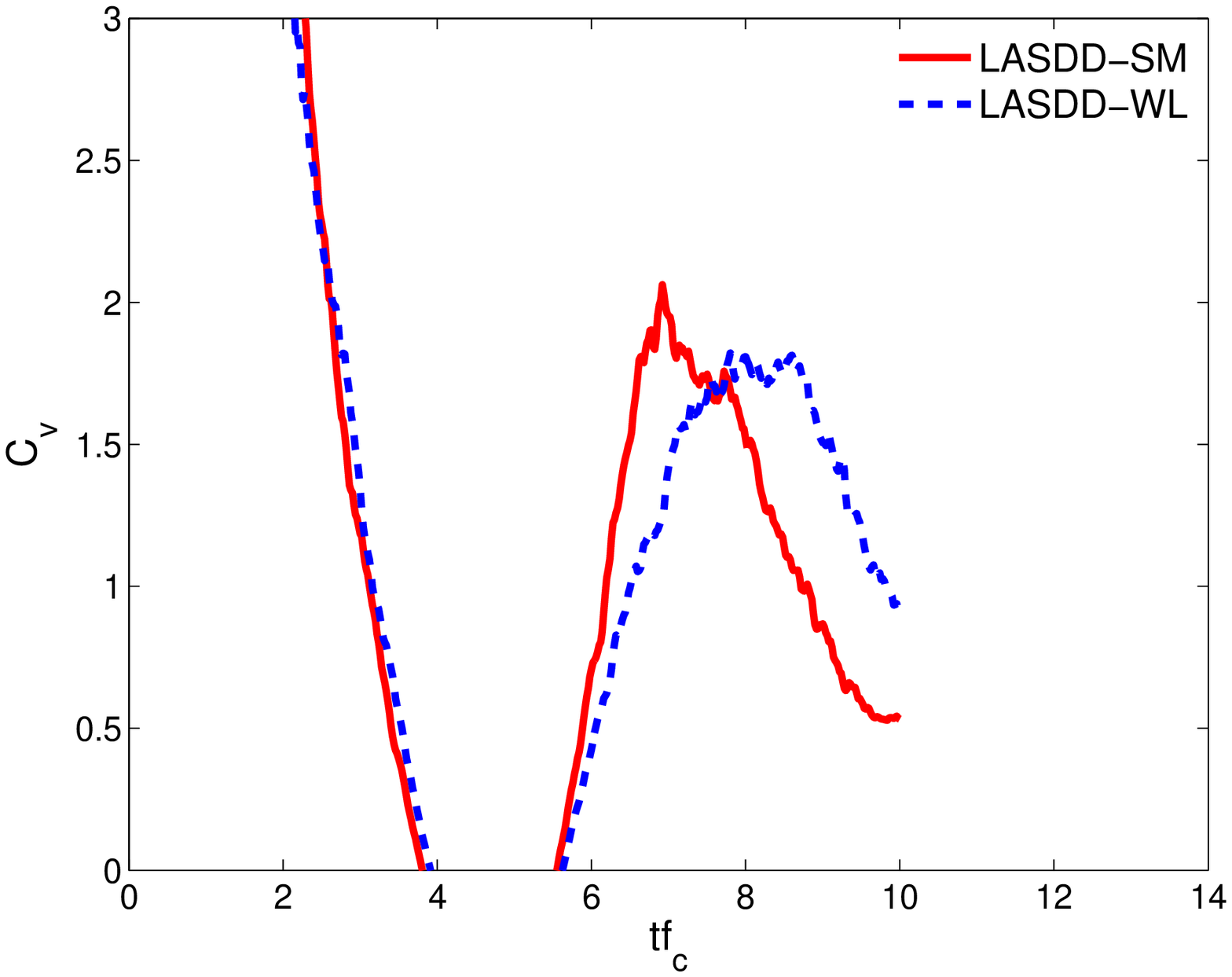}}
\caption{Temporal  evolution of  the nonstationarity  parameters $C_u$
(left) and $C_v$ (right).}
\label{Fig1}
\end{figure}

In Figure  \ref{Fig1}, we  present the nonstationary  parameters $C_u$
and $C_v$ (see
\cite{andr94} for  definitions). Under steady  state conditions,
these parameters should approach unity.  Although none of the past
\cite{andr94,chow05} and present simulations are quite close to steady
state conditions, they are more or  less in phase with each other. All
these simulations  clearly portray the inertial  oscillation of period
$2\pi/f_c$, as anticipated.

\begin{figure}[ht]
\centerline{\includegraphics[width=2.25in]{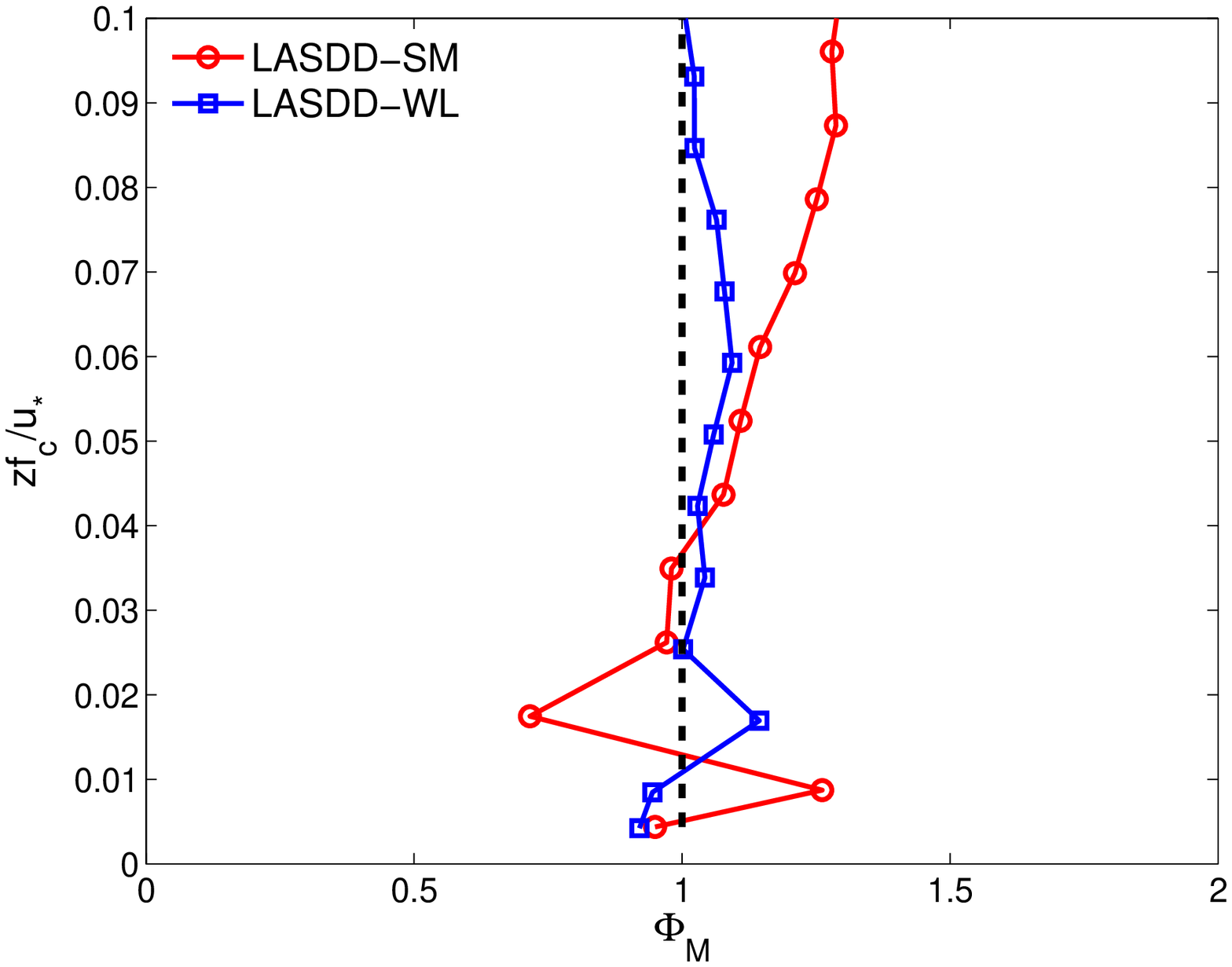}
\includegraphics[width=2.25in]{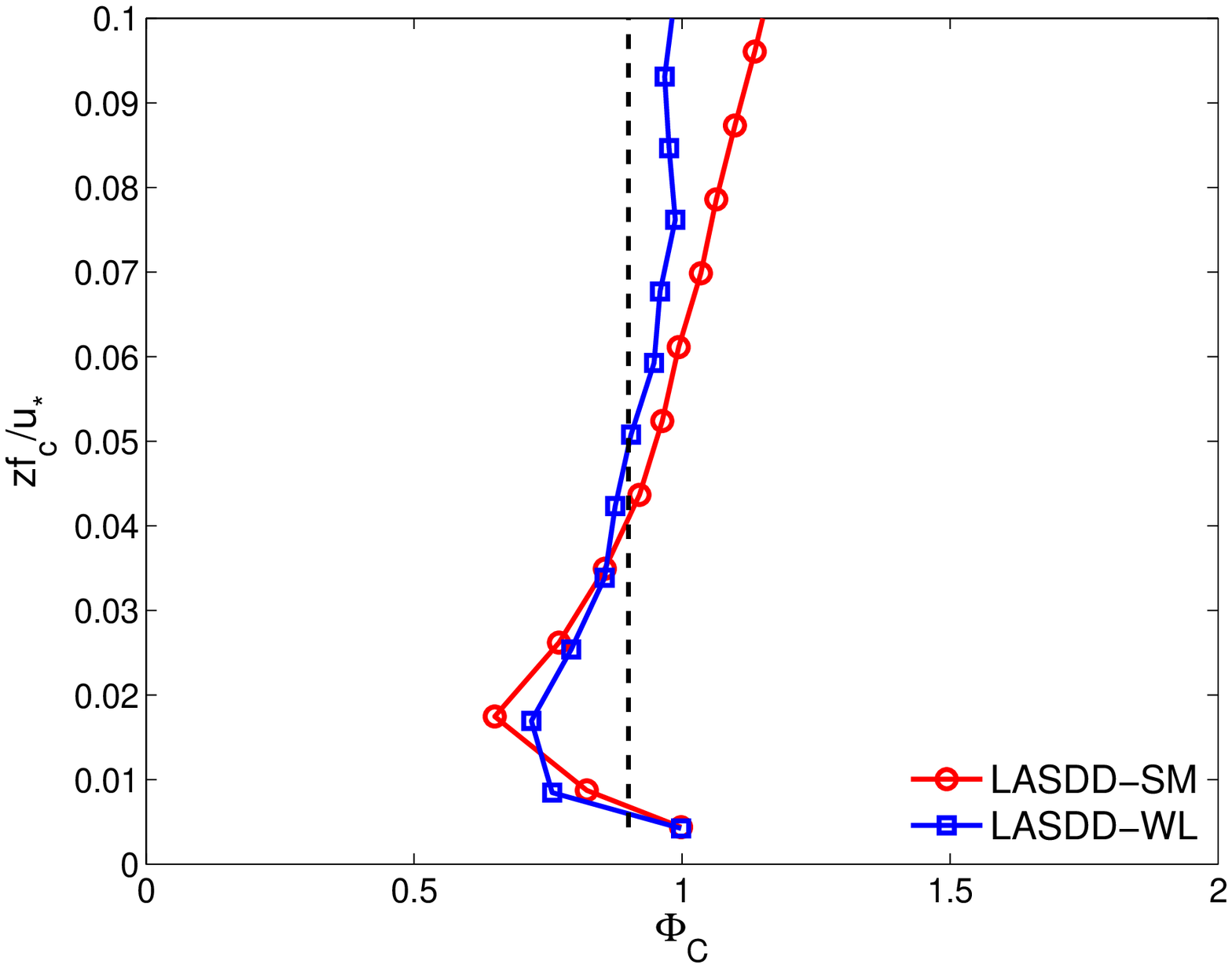}}
\caption{Simulated non-dimensional velocity  (left) and scalar (right)
gradients. The dashed  lines correspond to the values  of 1 (left) and
0.9 (right).  These  values are expected to hold  in the surface layer
under neutral conditions according to the similarity theory.}
\label{Fig2}
\end{figure}

Accurately   simulating    the   non-dimensional   velocity   gradient
$(\phi_M)$,  and  the  scalar  gradient $(\phi_C)$  in  the  neutrally
stratified surface layer has proven  to be a very challenging task for
many atmospheric  LES models.  It  is well known that  the traditional
Smagorinsky model is over-dissipative in the near-surface region and
gives rise to  excessive mean gradients in velocity  and scalar fields
(cf.  \cite{andr94}).   Fortunately, state-of-the-art LES-SGS modeling
approaches of Mason and Thomson \cite{maso92}, Sullivan et al.
\cite{sull94},   Kosovi\'{c}  \cite{koso97},  Port\'{e}-Agel   et  al.
\cite{port00}, Port\'{e}-Agel  \cite{port04}, Esau \cite{esau04}, Chow
et al.   \cite{chow05}, Bou-zeid et al.  \cite{bouz05},  and Stoll and
Port\'{e}-Agel \cite{stol06} offer major improvements over traditional
Smagorinsky-type  SGS   models,  and  reproduce   the  non-dimensional
gradients reasonably  well. From Figure  \ref{Fig2}, it is  clear that
both  the  LASDD-SM and  LASDD-WL  SGS  models behave  satisfactorily,
albeit, the  performance of  the LASDD-WL SGS  model is  superior.  We
would like  to stress that both  the LASDD SGS  modeling approaches do
not require any additional  stochastic term or supplementary near-wall
stress models for reliable performance in an LES.  In the framework of
Monin-Obukhov similarity theory, the non-dimensional velocity gradient
$(\phi_M)$ is indisputably equal to one (the dotted line in Figure
\ref{Fig2} - left).  However, in the literature there  is no consensus
on the `true' magnitude of the non-dimensional scalar gradient $(\phi_C)$. Businger et
al.  \cite{busi71},  based on the Kansas field  experiment, proposed a
value  of 0.74.   Recent field  observations, however,  suggest values
close  to 0.9  (for a  review, see  \cite{kade90}).  From  the present
coarse-resolution simulations,  it is difficult to favor either of these
values.   However, qualitatively,  both the  LASDD SGS  models portray
very similar non-dimensional scalar gradient profiles (Figure
\ref{Fig2} - right).

\begin{figure}[ht]
\centerline{\includegraphics[width=2.25in]{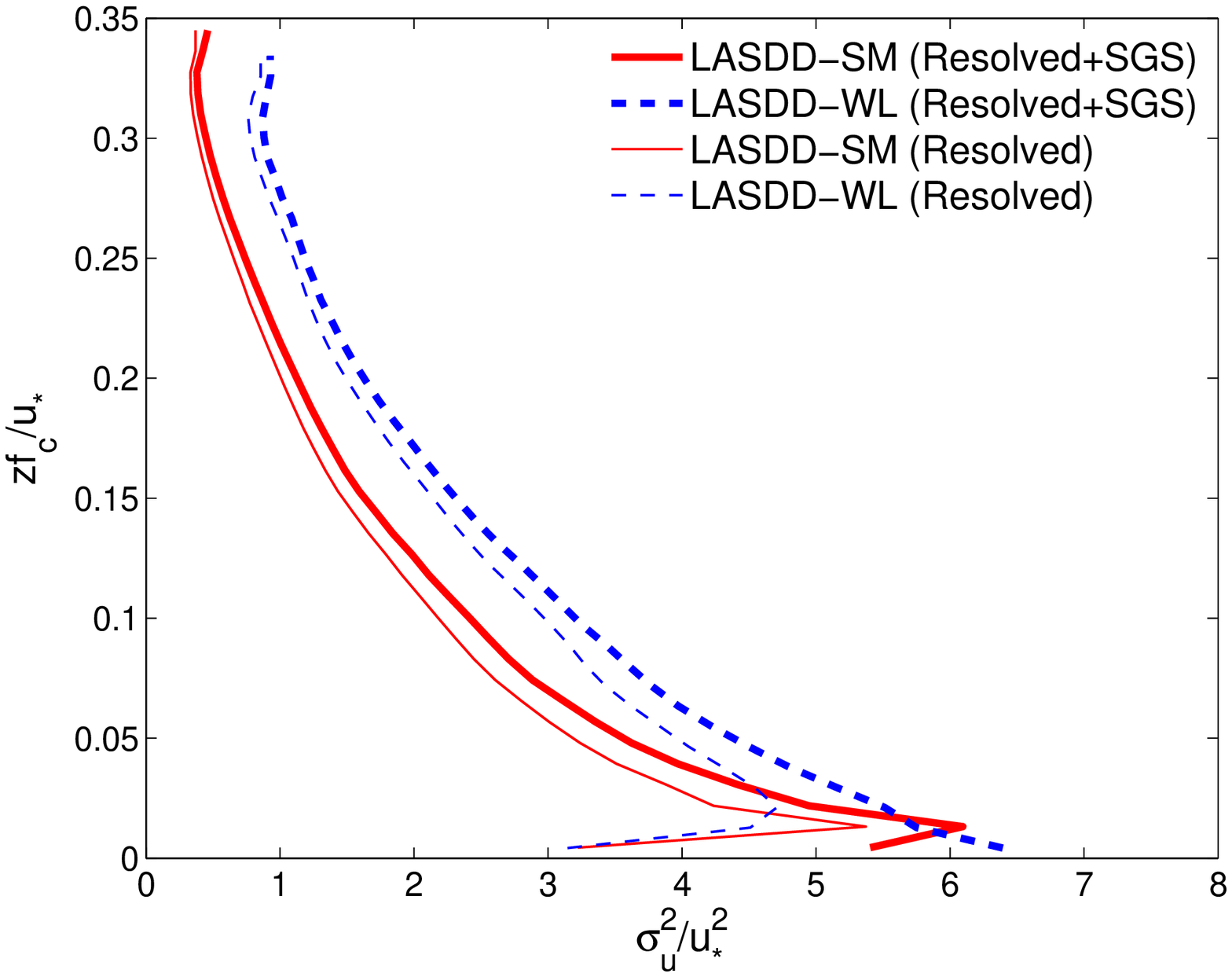}
\includegraphics[width=2.25in]{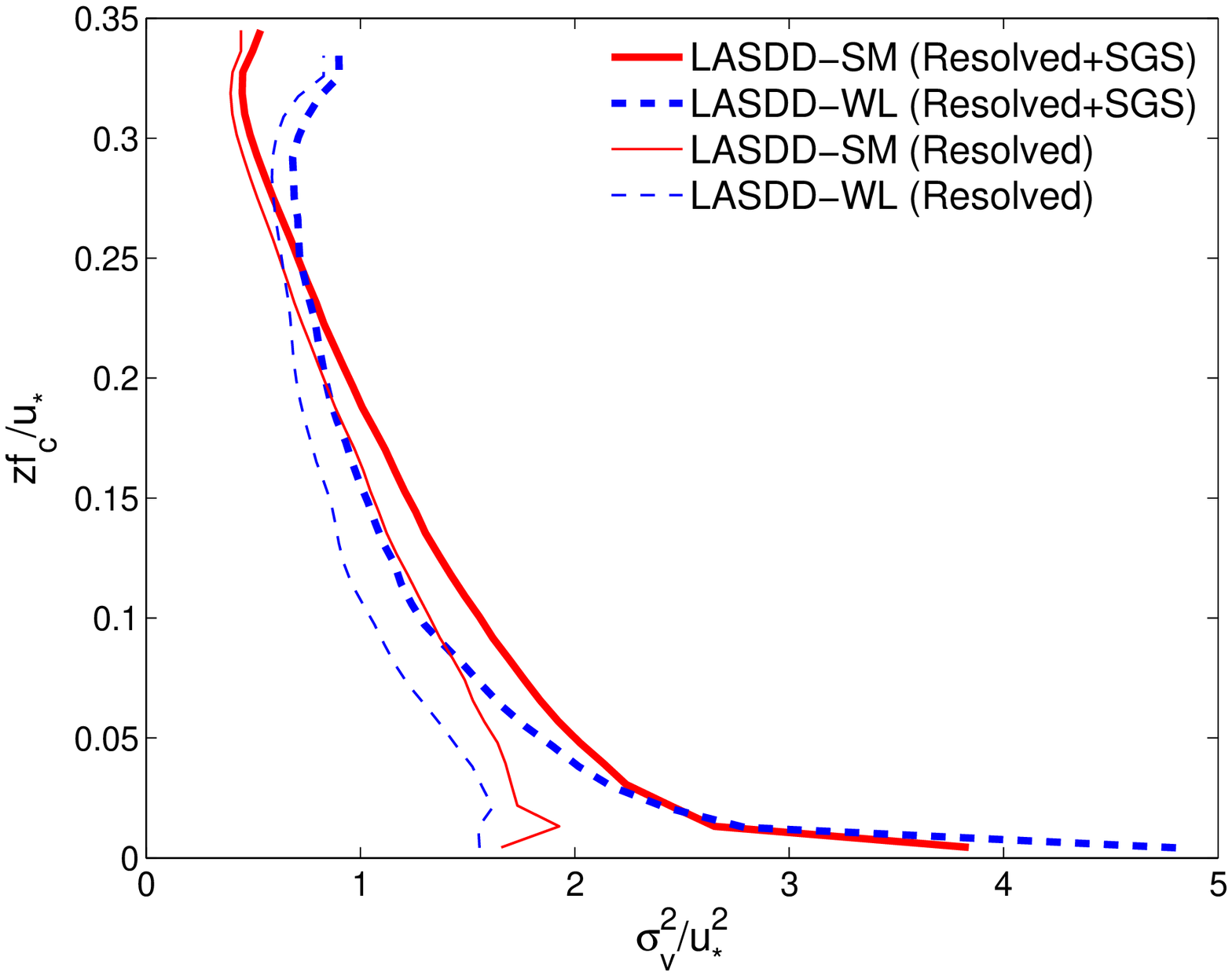}}
\centerline{\includegraphics[width=2.25in]{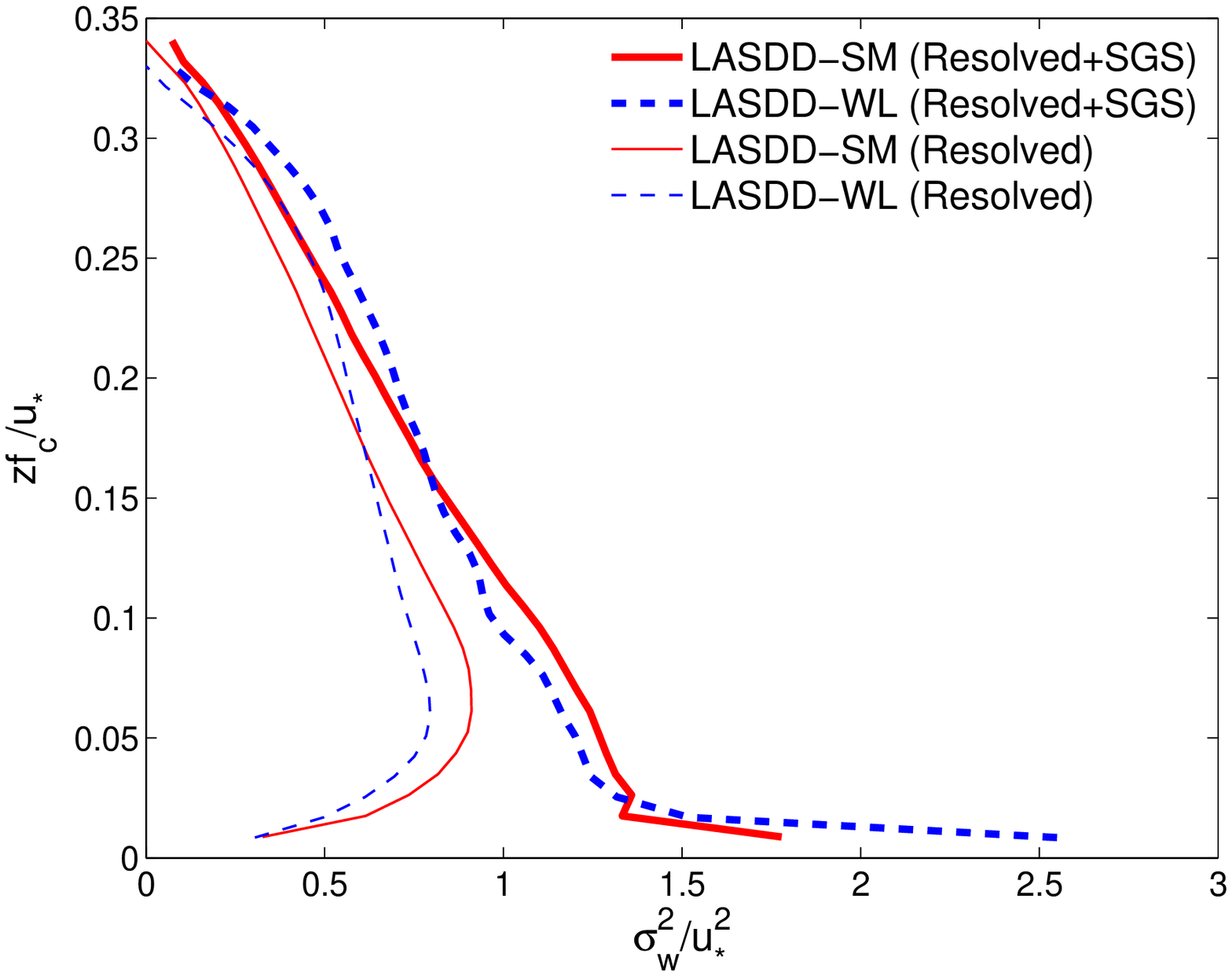}
\includegraphics[width=2.25in]{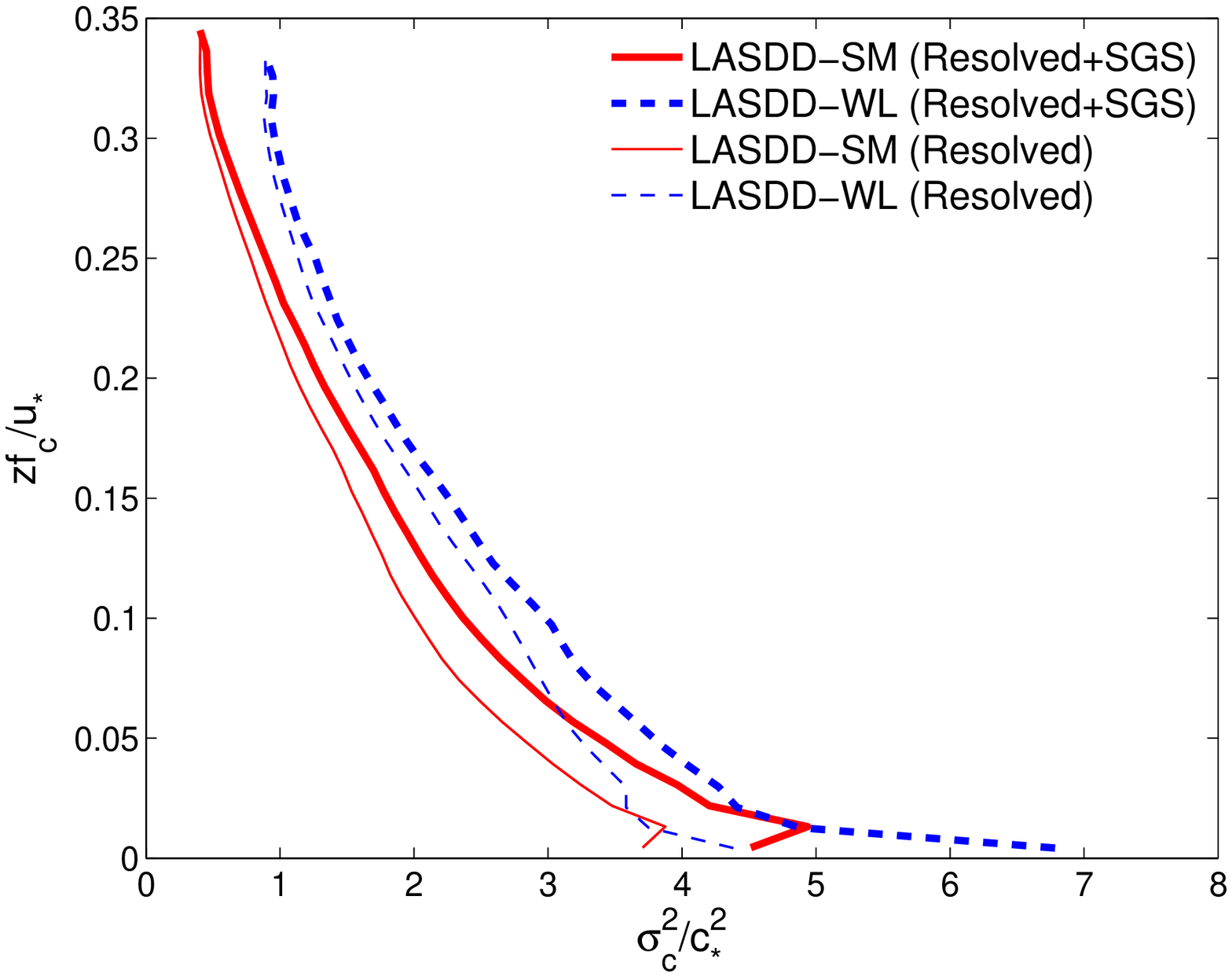}}
\caption{Simulated   normalized  longitudinal   (top-left),
transverse     (top-right),     vertical    (bottom-left)     velocity
variances. Simulated normalized scalar variances are shown in
the bottom-right plot.}
\label{Fig3}
\end{figure}

In  neutrally  stratified  ABL  flows, the  observed  peak  normalized
velocity variances  occur near the  surface and are of  the magnitude:
$\sigma_u^2/u_*^2   \sim  5-7$,   $\sigma_v^2/u_*^2  \sim   3-4,$  and
$\sigma_w^2/u_*^2 \sim  1-2$ \cite{gran92}.  The  corresponding values
found  in our  simulations (Figure  \ref{Fig3}) approximately  fall in
these  ranges.   The simulated  results  also  concur  with the  outer
boundary layer observations.  For example, the KONTUR data
\cite{gran86} give $\sigma_u^2/u_*^2 \sim \sigma_v^2/u_*^2 \sim 1$ and
$\sigma_w^2/u_*^2 \sim 0.5$ at $z  = 0.75z_i$ (where $z_i$ denotes the
inversion     height).      The     normalized    scalar     variances
$(\sigma_c^2/c_*^2)$ are also shown  in Figure \ref{Fig3}. Here, $c_*$
is the surface scalar scale ($= -\overline{wc}_0/u_*$). In
\cite{andr94}, it  was found that the  consensus among different
SGS models  is poorer in the  case of passive scalar  in comparison to
the  momentum case.   The disagreements  between different  SGS models
could   be  partially   attributed   to  different   {\it  a   priori}
prescriptions for the SGS  Prandtl ($Pr_{SGS}$) number, and underscore
the  need for  the determination  of $Pr_{SGS}$  in  a self-consistent
manner, as  is done in the  present study.  One  must also acknowledge
the  facts that  the passive  scalars exhibit  complex spatio-temporal
structure,  and the  statistical and  dynamical properties  of passive
scalars are  remarkably different from the  underlying velocity fields
\cite{warh00,shra00}.

We point out that the individual plots in Figure \ref{Fig3} represent 
both the normalized resolved and total (resolved + SGS) variances. In  the LASDD modeling
approach, one  does not solve additional prognostic  equations for the
SGS   turbulence   kinetic   energy   (TKE)   and   the   SGS   scalar
variances. However,  the SGS variances can be  roughly diagnosed using
the approach  of Mason \cite{maso89}. 

\begin{figure}[ht]
\centerline{\includegraphics[width=2.25in]{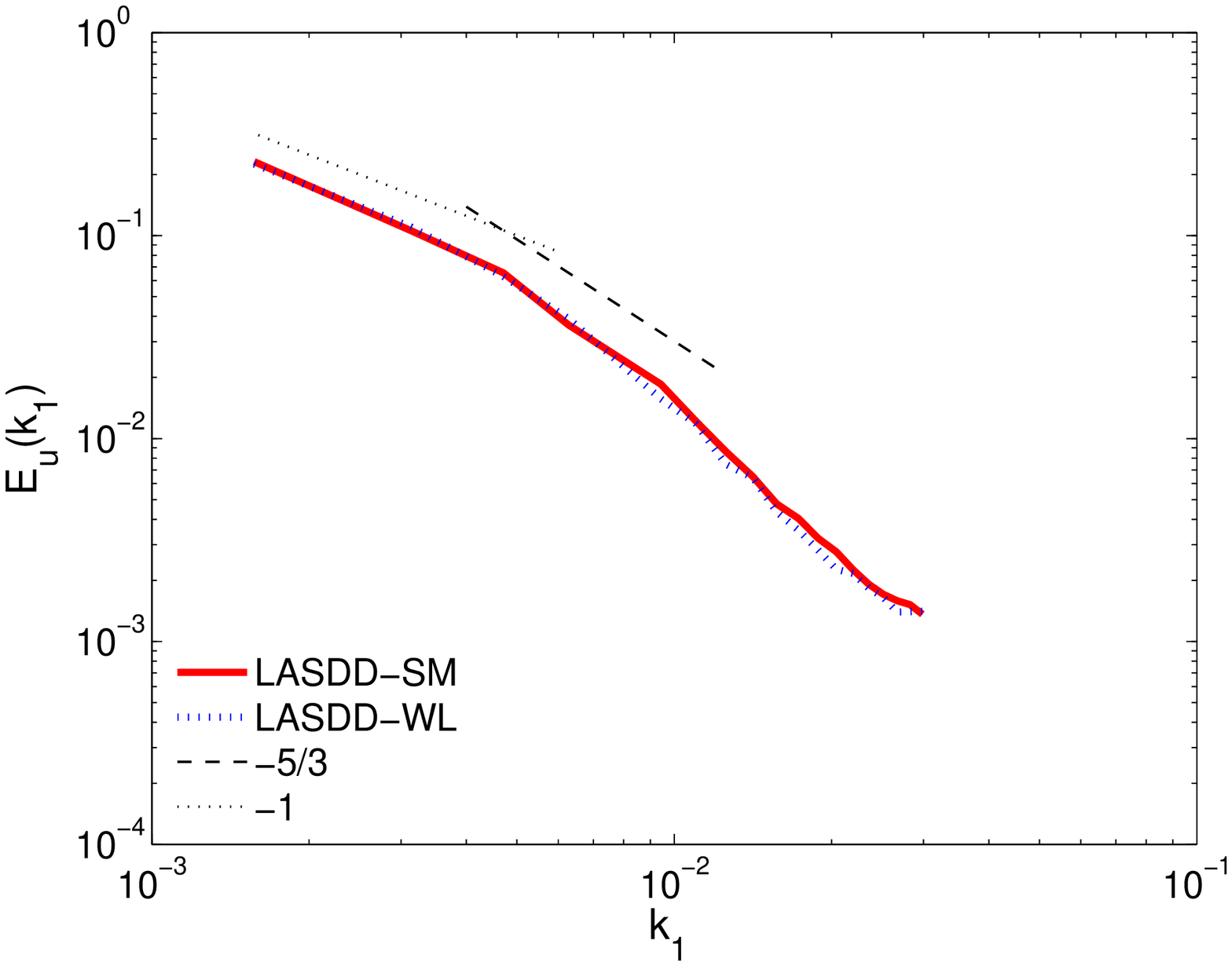}
\includegraphics[width=2.25in]{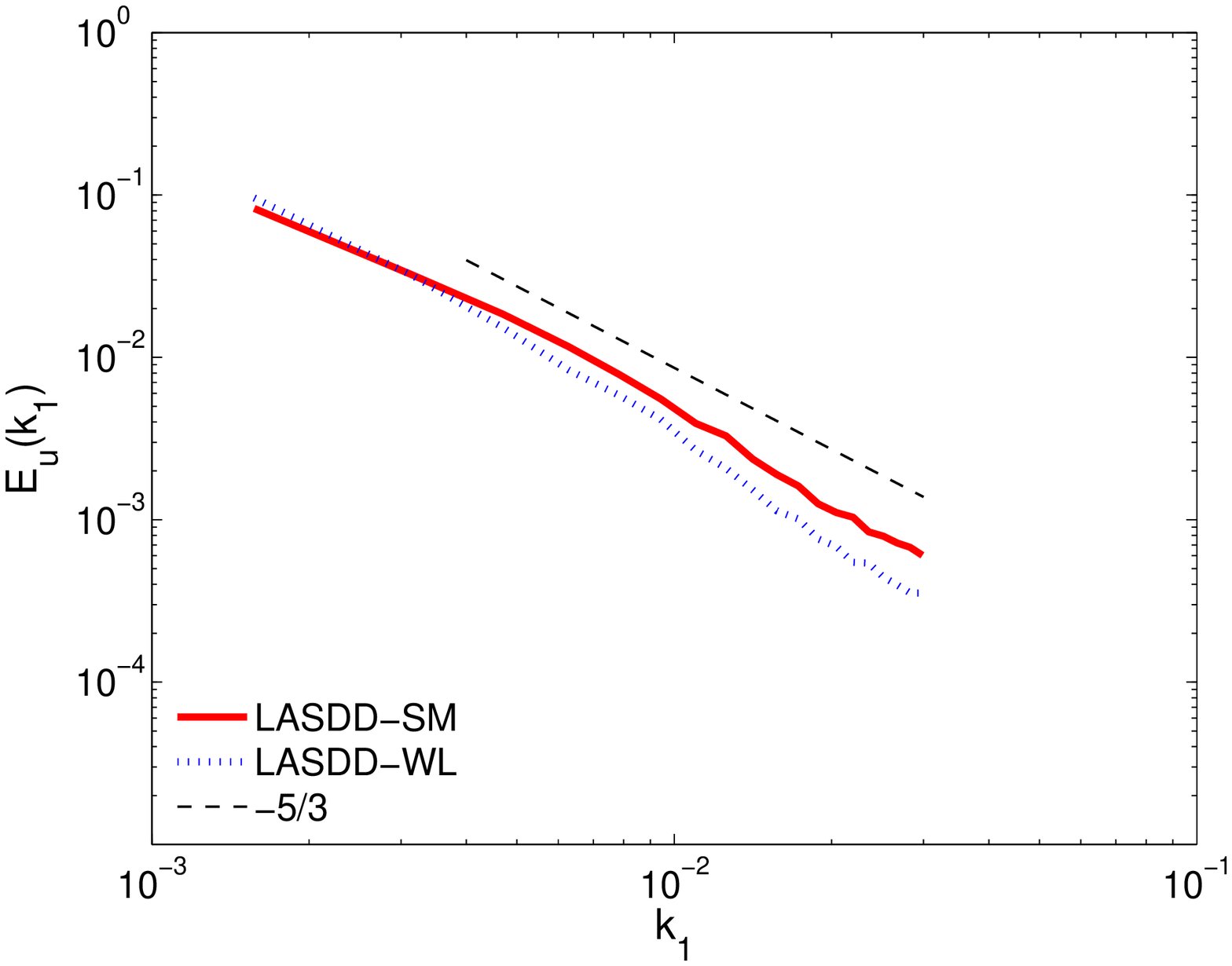}}
\centerline{\includegraphics[width=2.25in]{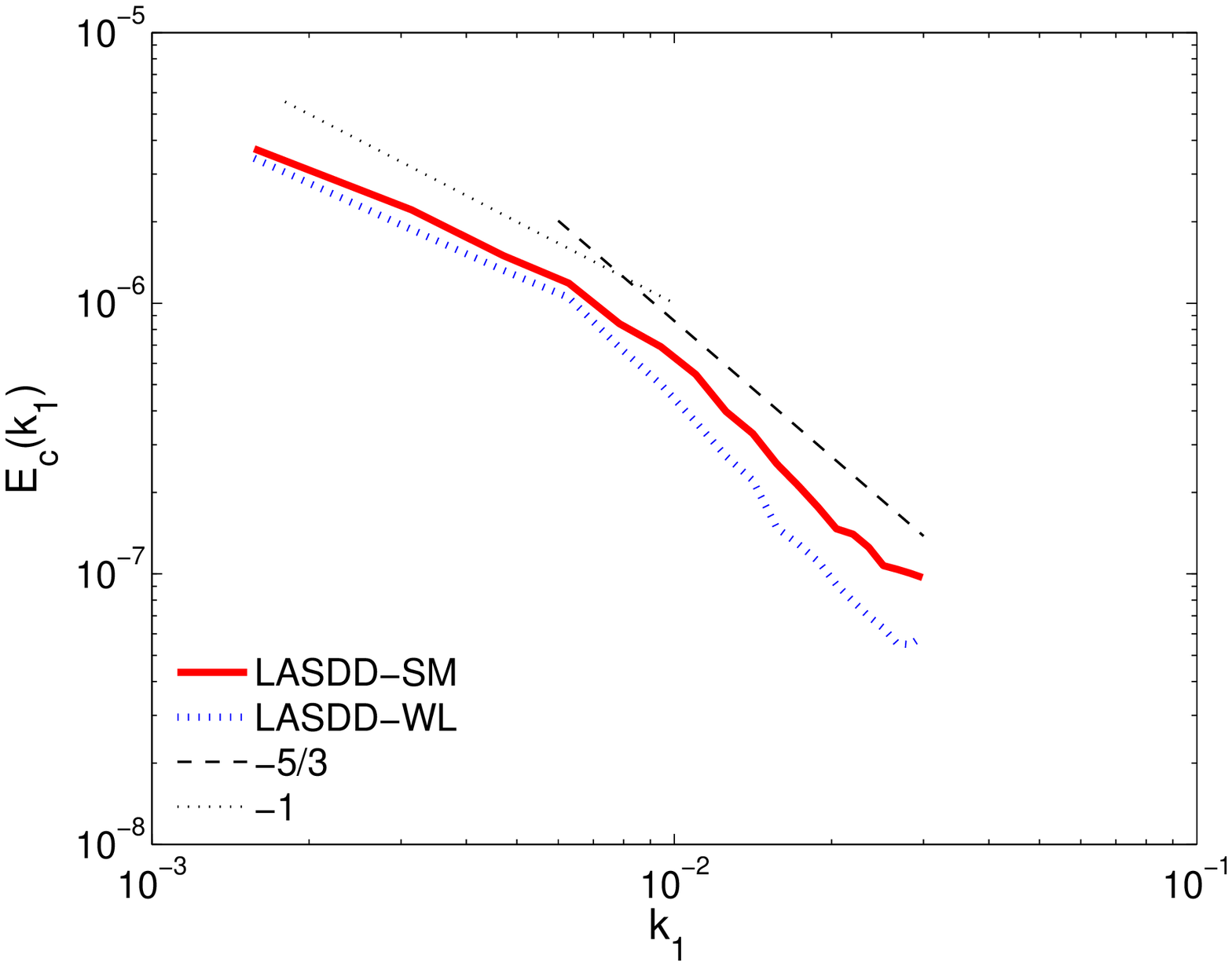}
\includegraphics[width=2.25in]{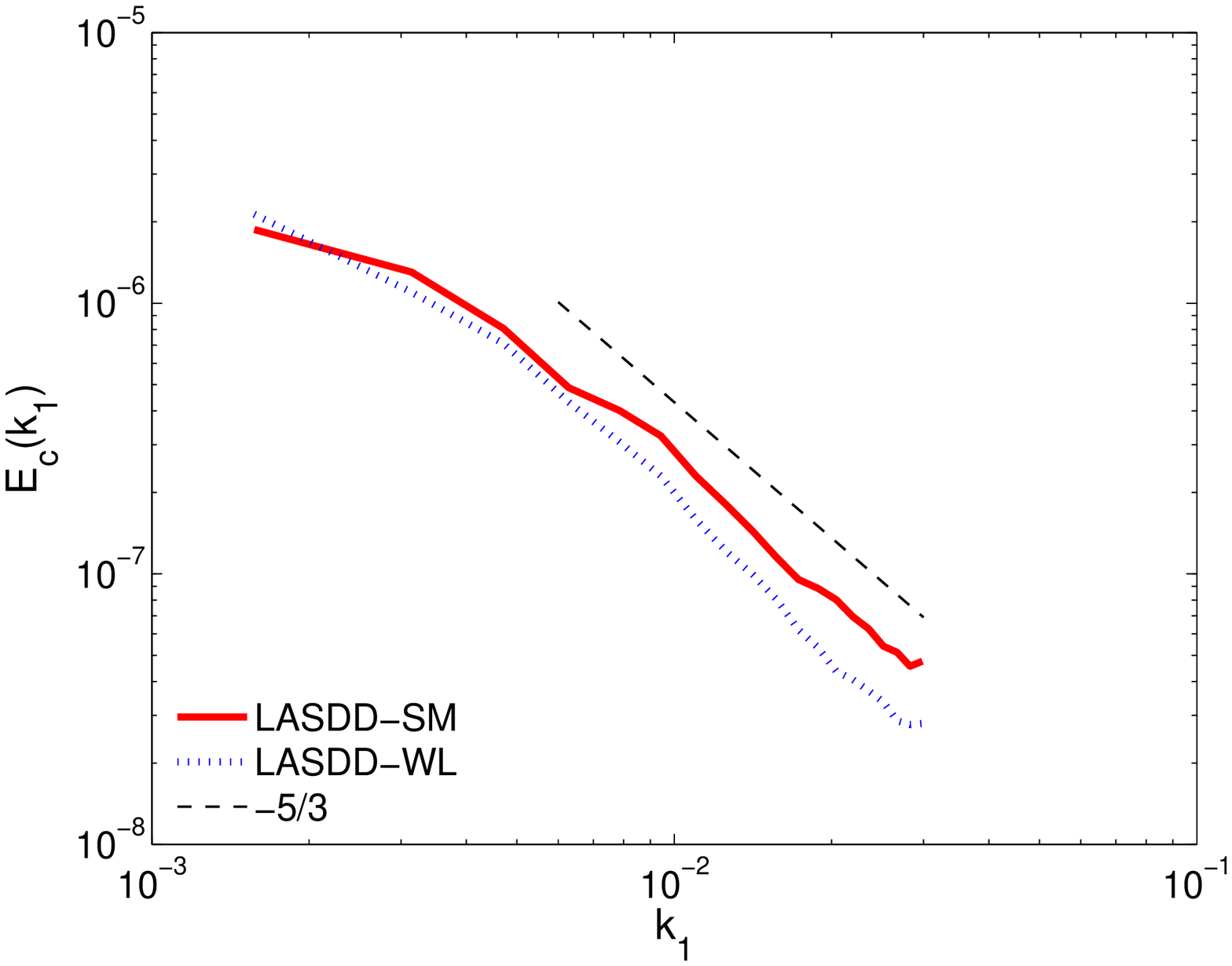}}
\caption{Spectra  of longitudinal velocity  (top), and  passive scalar
(bottom).   The  spectra  are   plotted  for  $z=0.1z_i$  (left),  and
$z=0.5z_i$ (right) levels. The dashed  and dotted  lines depict  the inertial
range  $(k_1^{-5/3})$  and  production  range  $(k_1^{-1})$  scalings,
respectively.}
\label{Fig4}
\end{figure}

The one-dimensional  longitudinal velocity and  passive scalar spectra
are computed  at heights $z=0.1z_i$, and $z=0.5z_i$,  and presented in
Figure \ref{Fig4}. The spectra highlight the most important difference
between the LASDD-SM  and LASDD-WL SGS models: the  LASDD-WL SGS model
seems to be over-dissipative  (indicated by steeper spectral slopes at
higher  wavenumbers).   In  the   case  of  the  LASDD-SM  model,  the
longitudinal  velocity  and   scalar  spectra  clearly  show  extended
inertial range ($k_1^{-5/3}$ scaling) at $z=0.5z_i$ . Near the surface
($z=0.1z_i$), the  longitudinal velocity spectra  show the anticipated
production  range ($k_1^{-1}$),  as well  as a  short  inertial range.
Recent research suggests that the production range is (likely) related
to elongated streaky velocity structures (see below).  Traditional SGS
models  typically do  not reproduce  well defined  inertial  ranges in
coarse-resolution   simulations  (cf.    \cite{andr94}).    From  that
perspective, the performance of the LASDD models could be considered a
near success.  We note  that the original plane-averaged \cite{port00}
and   the  Lagrangian-averaged   \cite{bouz05,stol06}  scale-dependent
dynamic  SGS  models  also   reproduced  the  characteristics  of  the
one-dimensional     longitudinal    velocity     spectra    remarkably
well. However, near the  surface, the passive scalar spectra predicted
by these SGS models showed unphysical pile up of scalar variances
\cite{port04,stol06}.   This  was possibly  due  to small  dynamically
determined eddy-diffusion coefficients near the surface
\cite{port04,stol06}. In  the present study we did  not encounter this
issue.

\begin{figure}[ht]
\centerline{\includegraphics[width=2.25in]{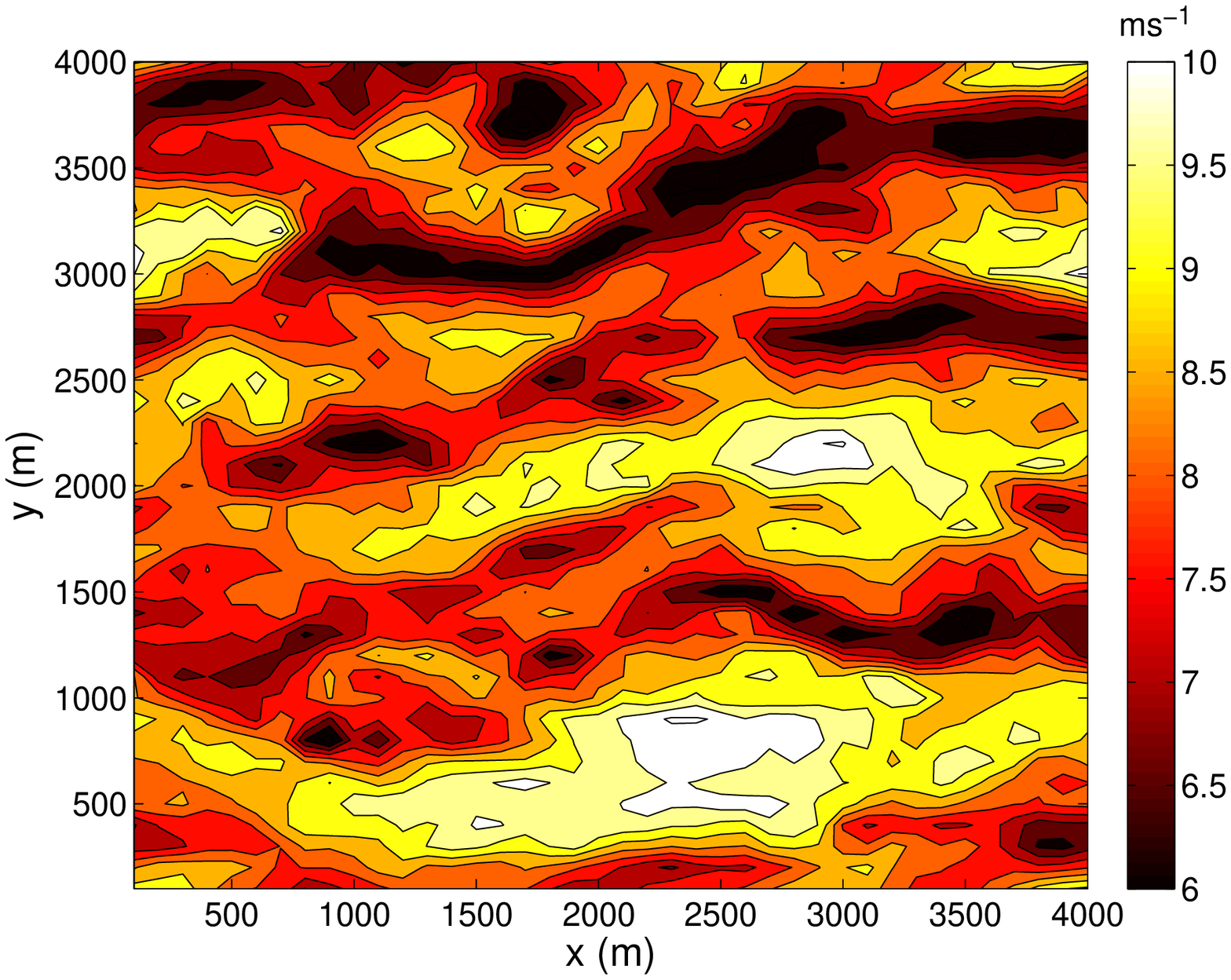}
\includegraphics[width=2.25in]{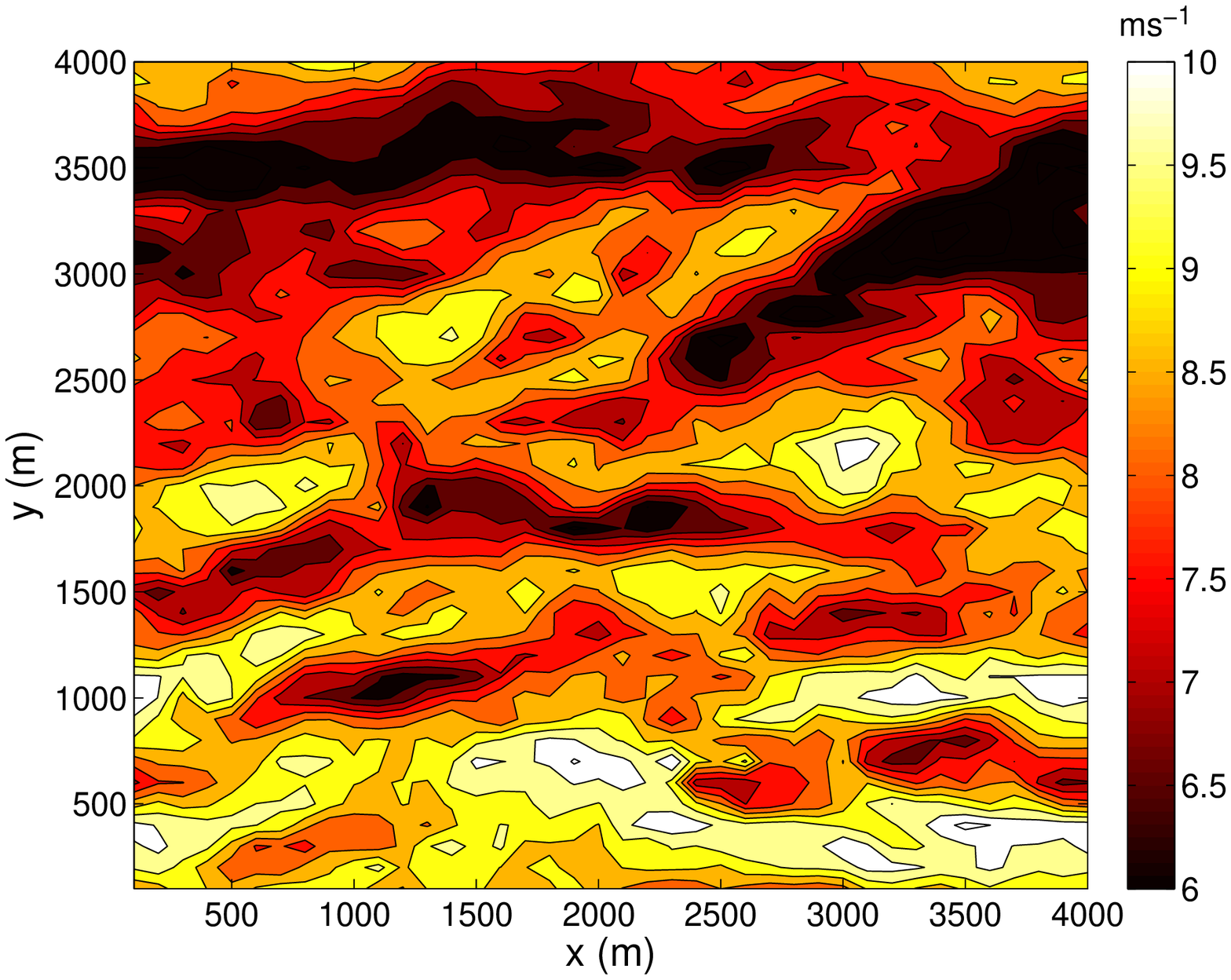}}
\centerline{\includegraphics[width=2.25in]{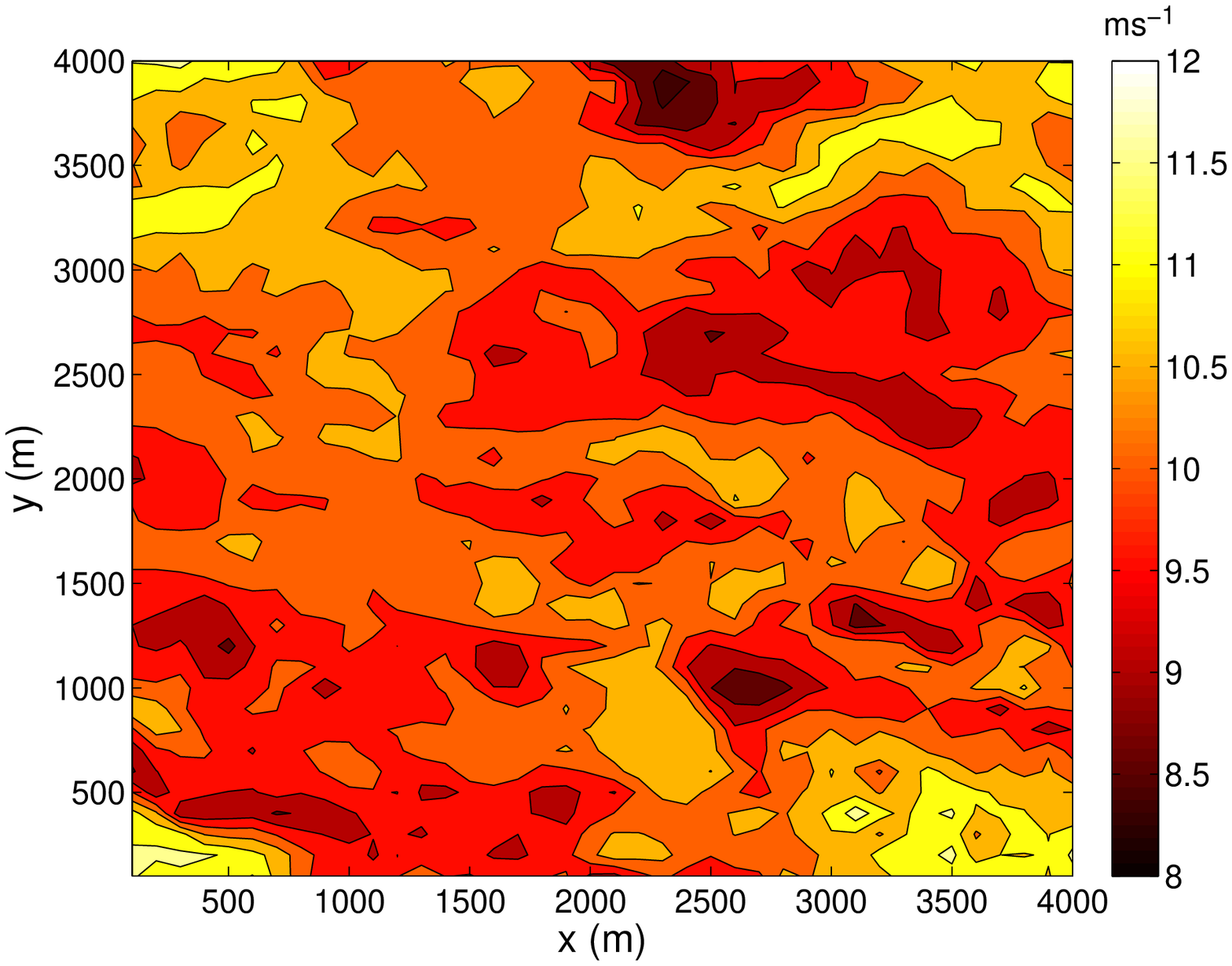}
\includegraphics[width=2.25in]{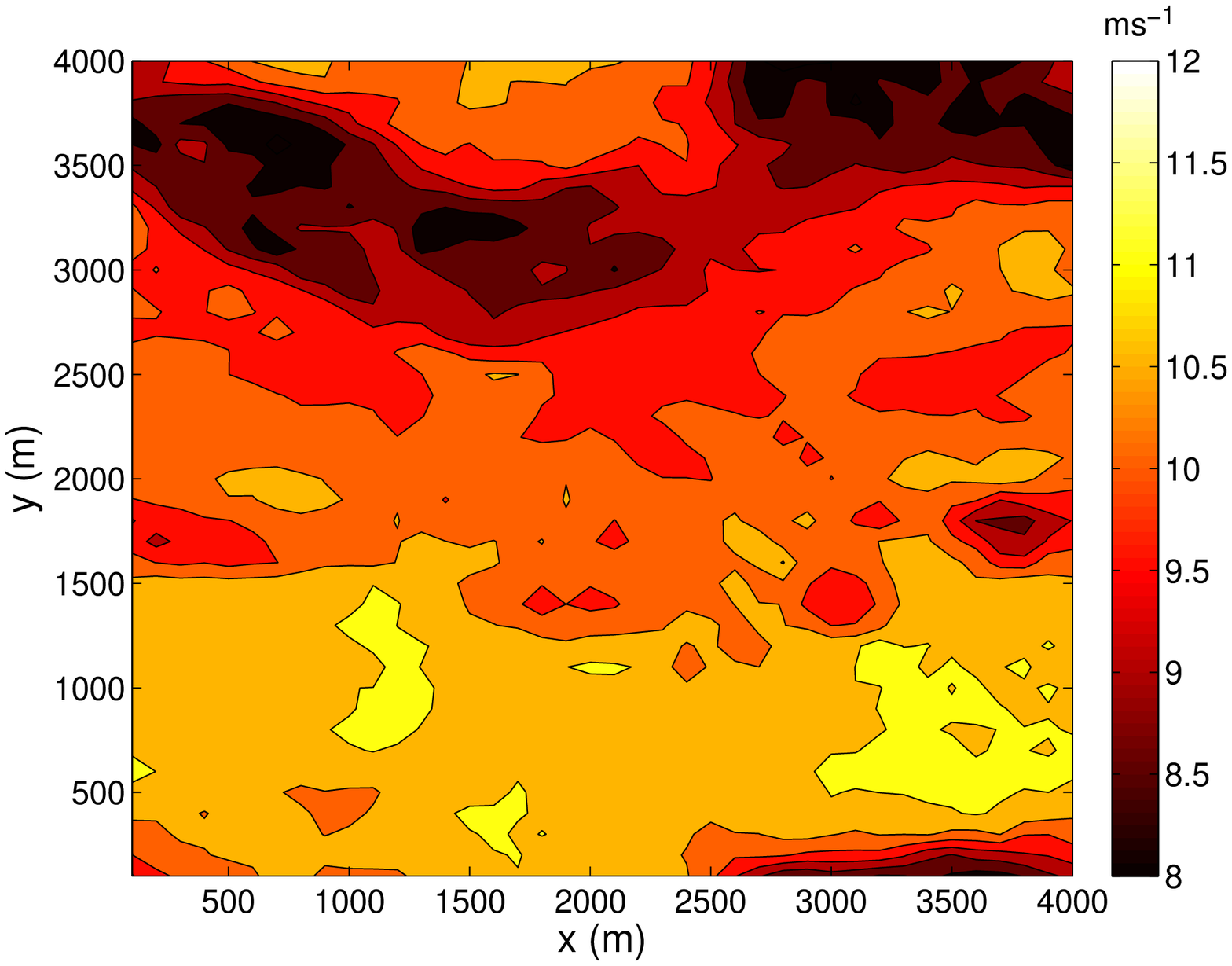}}
\caption{Visualization  of longitudinal  velocity fields  simulated by
LASDD-SM  (left),  and LASDD-WL  (right)  SGS  models. The  horizontal
cross-sections   are  taken  at   $z=0.1z_i$  (top),   and  $z=0.5z_i$
(bottom).}
\label{Fig5}
\end{figure}

A few  previous LES studies  have reported the existence  of elongated
streaky structures in the neutral surface layers
\cite{koso97,sull94,maso87,moen94,ding01,carl02}.   The  link  between
experimentally  observed long production  range ($k^{-1}$  scaling) in
the streamwise spectra of  the longitudinal velocity and the elongated
streaky structures has recently been discussed in depth by Carlotti
\cite{carl02}.   Moreover, strong  correlations between  these streaky
structures and large negative momentum flux were earlier reported by
\cite{moen94}. From Figure \ref{Fig5} (top), it is clear that both the
LASDD  models show streaky  structures, roughly  parallel to  the mean
wind  direction,  in  the  surface layer  (at  $z=0.1z_i$).   However,
significant morphologic differences are noticeable in the mid-ABL flow
structures.  In accordance with  past studies (cf. \cite{moen94}), the
LASDD-SM SGS model predicts  non-coherent structures at $z=0.5z_i$. In
contrast,  large coherent  structures  persist in  the LASDD-WL  model
results   (Figure  \ref{Fig5},  bottom-right).    Another  interesting
feature of this plot is the (virtual) non-existence of fine-scale flow
structures. This can be  directly associated with the over-dissipative
nature of the LASDD-WL SGS  model, as discussed before. In essence, we
can infer that the (non-)existence of coherent structures in NBL flows
are  strongly  dependent  on  SGS  parameterizations,  especially  for
coarse-resolution  simulations.   A   few  previous  studies  somewhat
support this inference.  For instance, the nonlinear SGS model
\cite{koso97},  and the modified  Smagorinsky SGS  model \cite{ding01}
barely produced any elongated streaky structures.

\section{Concluding Remarks}\label{Sec4}

Two  locally  averaged scale-dependent  dynamic  SGS  closures --  the
LASDD-SM \cite{basu06a,basu06b} and the  LASDD-WL (this study) -- have
been used  to simulate a  neutral ABL case.  Although  the theoretical
foundations of  these SGS models are  fundamentally different, results
presented in  Figures \ref{Fig1} through  \ref{Fig5} illustrate strong
congruence between their results,  and with firmly established results
(i.e.   the Monin-Obukhov  similarity  theory and  the inertial  range
scaling  of  spectra).   The  normalized  variances  computed  in  our
simulations  also  closely  follow  the  ones  calculated  from  field
measurements.  The major  noticeable and consistent difference between
the results is shown in Figure
\ref{Fig4}: the  LASDD-WL SGS model appears to  be over-dissipative at
the higher  wavenumbers, in comparison  to the LASDD-SM SGS  model. In
Figure  \ref{Fig5}, we  see  that both  SGS  models predict  elongated
streaky  structures  in  the  near-wall  region  ($z=0.1z_i$).   These
coherent structures are  no longer evident at higher  locations in the
domain, in the  case of the LASDD-SM SGS  model-based simulation.  Due
to undue dissipation, the  LASDD-WL SGS model-based simulation results
in quite different flow structures at this level.

The Wong-Lilly SGS base model requires fewer assumptions and comes at slightly less
computational cost in comparison to the commonly used Smagorinsky 
SGS base model.  Unfortunately,  these advantages seem to be  offset by its
over-dissipative   tendency at higher wavenumbers.   Some   inherent   assumptions  of   the
Smagorinsky base model can also  be eliminated by solving a prognostic
equation for the TKE. However, when using  this TKE SGS approach, the SGS model
coefficients are often tuned for different ABL flow conditions
\cite{sull94,sull03}. An alternative approach would be to formulate a
dynamic  version of the  TKE SGS  model, which  will also  account for
energy backscatter.  We are currently  working on this SGS approach to
better represent the physics of atmospheric boundary layer flows.

\acknowledgement{This work was partially funded by the National Institute 
of Standards and Technology, the National Science  Foundation and the  Texas 
Advanced  Research Program grants.   All the  computational resources
were kindly provided  by the High Performance Computing  Center at Texas 
Tech University.}

\section*{Appendix}
\setcounter{equation}{0}
\renewcommand{\theequation}{A\arabic{equation}}

The SGS model proposed by  Wong and Lilly \cite{wong94} can be written
as:
\begin{equation}
\tau_{ij} - \frac{1}{3}\tau_{kk}\delta_{ij} = -2C_{WL}\Delta_f^{4/3}\tilde{S}_{ij}
\label{Eq1}
\end{equation} 
where $\tau_{ij}$  and $\tilde{S}_{ij}$  denote the SGS  stress tensor
and the resolved strain rate tensor, respectively. $C_{WL}$ is a model
coefficient to  be specified or  determined dynamically.  In  a recent
LES study of  neutral boundary layer flows, Chow  et al. \cite{chow05}
utilized a dynamic  version of this SGS model  in conjunction with the
approximate deconvolution  model (ADM) for  resolvable subfilter-scale
(RSFS) components.  To account for the smaller underresolved eddies in
the surface layer,  they used a near-wall stress  model in addition to
the  dynamic Wong-Lilly SGS  and ADM-RSFS  models.  As  an alternative
approach,   in   this   work,   we  formulate   a   locally   averaged
scale-dependent  dynamic   version  of  Equation   (\ref{Eq1})  (named
LASDD-WL).

The SGS  stress tensor ($\tau_{ij}$) at the  filter scale ($\Delta_f$)
is defined as: $\tau_{ij} = \widetilde{u_i u_j} - \widetilde{u_i}
\widetilde{u_j}$.  In  a seminal work,  Germano et al. \cite{germ91} proposed
to invoke an additional explicit test filter of width $\alpha
\Delta_f$  in  order  to  dynamically compute  the  SGS  coefficients.
Consecutive filtering  at scales  $\Delta_f$ and at  $\alpha \Delta_f$
leads to a  SGS turbulent stress tensor ($T_{ij}$)  at the test filter
scale $\alpha \Delta_f$:
\begin{equation}
T_{ij}  = \overline{\widetilde{u_i~u_j}}  - \overline{\widetilde{u}_i}
~\overline{\widetilde{u}_j},
\end{equation}
where an  overline $\overline{(\cdots)}$ denotes filtering  at a scale
of  $\alpha  \Delta_f$.   From  the  definitions  of  $\tau_{ij}$  and
$T_{ij}$ an algebraic relation can  be formed, known in the literature
as the Germano identity:
\begin{equation}
L_{ij} = \overline{\widetilde{u}_i\widetilde{u}_j} -
\overline{\widetilde{u}_i}~\overline{\widetilde{u}_j}   =   T_{ij}   -
\overline{\tau_{ij}}.
\end{equation}
This identity  is then effectively used to  dynamically obtain unknown
SGS model coefficients.  In the case of the Wong-Lilly model (Equation
(\ref{Eq1})), this identity yields:
\begin{equation}
L_{ij} - \frac{1}{3}L_{kk}\delta_{ij} =
\left(C_{WL}\right)_{\Delta_f}M_{ij},
\end{equation}
where $M_{ij} = 2\Delta_f^{4/3}\left( 1 - \alpha^{4/3}
\frac {\left(C_{WL}\right)_{\alpha\Delta_f}}
{\left(C_{WL}\right)_{\Delta_f}} \right)
\overline{\widetilde{S_{ij}}}$.   If  one  assumes  scale  invariance,
i.e., $\left(C_{WL}\right)_{\alpha \Delta_f} =
\left(C_{WL}\right)_{\Delta_f}$,   then   the   unknown
coefficient $\left(C_{WL}\right)_{\Delta_f}$  can be easily determined
following the error minimization approach of Lilly \cite{lill92}:
\begin{equation}
\left(C_{WL}\right)_{\Delta_f}   =  \frac{\langle  L_{ij}M_{ij}\rangle}
{\langle M_{ij}M_{ij}\rangle}.
\label{EqLijMij}
\end{equation}
In   the  context  of   the  present   study,  the   angular  brackets
$\langle\cdots\rangle$   denote   localized   spatial   averaging   on
horizontal planes with a stencil of three by three grid points
\cite{basu06a,basu06b}.

Recent studies have  shown that the assumption of  scale invariance is
seriously flawed for sheared and stratified boundary layer flows
\cite{port00,port04,basu06a,basu06b,bouz05,stol06}. In  other words,
the     ratio     of     $\left(C_{WL}\right)_{\alpha\Delta_f}$     to
$\left(C_{WL}\right)_{\Delta_f}$  should not be  assumed equal  to one
for most  of these ABL flow scenarios.   Rather, this scale-dependence
ratio  should be determined  dynamically.  In  order to  implement the
scale-dependent dynamic  procedure, one needs to employ  a second test
filtering  operation at  a scale  of $\alpha^2  \Delta_f$  [denoted by
$\widehat{(\cdots)}$].  Invoking  the Germano identity  for the second
time leads to:
\begin{equation}
Q_{ij} - \frac{1}{3}Q_{kk}\delta_{ij} =
\left(C_{WL}\right)_{\Delta_f}N_{ij},
\label{EqQij}
\end{equation}
where
\begin{eqnarray*}
Q_{ij} = \widehat{\widetilde{u_i}\widetilde{u_j}} -
\widehat{\widetilde{u_i}}\widehat{\widetilde{u_j}} \nonumber
\end{eqnarray*}
and
\begin{eqnarray*}
N_{ij} = 2\Delta_f^{4/3}\left( 1 - \alpha^{8/3}
\frac {\left(C_{WL}\right)_{\alpha^2\Delta_f}}
{\left(C_{WL}\right)_{\Delta_f}} \right) \widehat{\widetilde{S_{ij}}}.
\nonumber
\end{eqnarray*} 
This results in:
\begin{equation}
\left(C_{WL}\right)_{\Delta_f}   =  \frac{\langle  Q_{ij}N_{ij}\rangle}
{\langle N_{ij}N_{ij}\rangle}.
\label{EqQijNij}
\end{equation} 
Following \cite{port00}, the following scale-dependence assumption can
be made:
\begin{equation}
\beta           =           \frac{\left(C_{WL}\right)_{\alpha\Delta_f}}
{\left(C_{WL}\right)_{\Delta_f}} =
\frac{\left(C_{WL}\right)_{\alpha^2\Delta_f}}
{\left(C_{WL}\right)_{\alpha\Delta_f}},
\label{beta}
\end{equation}
This is  a much weaker  assumption than the  scale-invariance modeling
assumption of  $\beta = 1$.  Now, from  Equations (\ref{EqLijMij}) and
(\ref{EqQijNij}),  using  Equation (\ref{beta}),  one  solves for  the
unknown  parameter $\beta$,  which  in  turn is  used  to compute  the
Wong-Lilly  SGS  model coefficient,  $\left(C_{WL}\right)_{\Delta_f}$,
utilizing Equation (\ref{EqLijMij}).

Solving  for  $\beta$ essentially  involves  finding  the  roots of  a
fifth-order polynomial \cite{port00}:
\begin{equation}
A_0 + A_1\beta + A_2\beta^2 + A_3\beta^3 + A_4\beta^4 + A_5\beta^5 = 0
\label{poly}
\end{equation}
where $A_0 = a_1a_3 - a_6a_8$,  $A_1 = a_1a_4 - a_7a_8$, $A_2 = a_2a_3
+  a_1a_5  -  a_6a_9$, $A_3  =  a_2a_4  -  a_7a_9$,  $A_4 =  a_2a_5  -
a_6a_{10}$, and $A_5 = -a_7a_{10}$. In the case of Wong-Lilly SGS base
model,         we        derive:         $a_1         =        \langle
Q_{ij}\widehat{\widetilde{S_{ij}}}\rangle$,     $a_2     =     \langle
-\alpha^{8/3}Q_{ij}\widehat{\widetilde{S_{ij}}}\rangle$, $a_3 =
\langle  {\overline{\widetilde{S_{ij}}}}^2  \rangle$,  $a_4 =  \langle
-2\alpha^{4/3} {\overline{\widetilde{S_{ij}}}}^2 \rangle$, $a_5 =
\langle \alpha^{8/3}  {\overline{\widetilde{S_{ij}}}}^2 \rangle$, $a_6
= \langle L_{ij}{\overline{\widetilde{S_{ij}}}} \rangle$, $a_7 =
\langle  -\alpha^{4/3}L_{ij}{\overline{\widetilde{S_{ij}}}}  \rangle$,
$a_8 = \langle {\widehat{\widetilde{S_{ij}}}}^2 \rangle$, $a_9 =
\langle  -2\alpha^{8/3}{\widehat{\widetilde{S_{ij}}}}^2  \rangle$, and
$a_{10} = \langle \alpha^{16/3} {\widehat{\widetilde{S_{ij}}}}^2
\rangle$.   Please  note that  the  coefficients  ($a_1$ to  $a_{10}$)
involve significantly  lesser number of tensor terms  in comparison to
the  ones derived by  Port\'{e}-Agel et  al.  \cite{port00}  using the
Smagorinsky   SGS base model.    Lesser   number  of   calculations
(specifically the tensor  multiplications) undoubtedly lead to cheaper
computational costs.

Scale-dependent formulation  for scalars can  be derived in  a similar
manner \cite{port04}. The Wong-Lilly  model for a generic scalar ($c$)
could be written as:
\begin{equation}
q_{i}                                                                 =
-\frac{C_{WL}}{Pr_{SGS}}\Delta_f^{4/3}\frac{\partial\tilde{c}}{\partial
x_i}
\label{EqC}
\end{equation} 
where $Pr_{SGS}$ is the so-called  SGS Prandtl number.  In the dynamic
or scale-dependent  dynamic modeling approaches,  typically the lumped
SGS   coefficient   ($C_{WL}Pr_{SGS}^{-1}$)   is   determined   in   a
self-consistent manner.   This procedure not only  eliminates the need
for any ad  hoc assumption about the SGS  Prandtl number ($Pr_{SGS}$),
it also completely  decouples the SGS scalar flux  estimation from SGS
stress computation. In the scale-dependent approach \cite{port04}, one
further defines  a scale-dependent parameter  for scalars ($\beta_c$),
analogous to Equation (\ref{beta}). For the Wong-Lilly SGS base model,
it could be written as:
\begin{equation}
\beta_c           =           \frac{\left(C_{WL}Pr_{SGS}^{-1}\right)_{\alpha\Delta_f}}
{\left(C_{WL}Pr_{SGS}^{-1}\right)_{\Delta_f}} =
\frac{\left(C_{WL}Pr_{SGS}^{-1}\right)_{\alpha^2\Delta_f}}
{\left(C_{WL}Pr_{SGS}^{-1}\right)_{\alpha\Delta_f}},
\label{betaC}
\end{equation}
As before,  $\beta_c$ could be  determined by solving  the fifth-order
polynomial:
\begin{equation}
A_0  +  A_1\beta_c +  A_2\beta_c^2  +  A_3\beta_c^3  + A_4\beta_c^4  +
A_5\beta_c^5 = 0
\label{polyC}
\end{equation}
where $A_0 = a_1a_3 - a_6a_8$,  $A_1 = a_1a_4 - a_7a_8$, $A_2 = a_2a_3
+  a_1a_5  -  a_6a_9$, $A_3  =  a_2a_4  -  a_7a_9$,  $A_4 =  a_2a_5  -
a_6a_{10}$, and $A_5 = -a_7a_{10}$.  For the Wong-Lilly SGS base model
for scalars, we get: $a_1 = \langle K_i'\widehat{\frac{\partial
\widetilde{c}}{\partial  x_i}} \rangle$,  $a_2  = \langle-\alpha^{8/3}
K_i'\widehat{\frac{\partial  \widetilde{c}}{\partial  x_i}}  \rangle$,
$a_3  =  \langle  {\overline  {\frac{\partial  \widetilde{c}}{\partial
x_i}}}^2   \rangle$,   $a_4   =  \langle   -2\alpha^{4/3}   {\overline
{\frac{\partial \widetilde{c}}{\partial x_i}}}^2 \rangle$, $a_5 =
\langle         \alpha^{8/3}         {\overline        {\frac{\partial
\widetilde{c}}{\partial    x_i}}}^2   \rangle$,    $a_6    =   \langle
K_i\overline{\frac{\partial  \widetilde{c}}{\partial  x_i}}  \rangle$,
$a_7 = \langle - \alpha^{4/3} K_i\overline{\frac{\partial
\widetilde{c}}{\partial  x_i}}  \rangle$,  $a_8  =  \langle  {\widehat
{\frac{\partial \widetilde{c}}{\partial x_i}}}^2 \rangle$, $a_9 =
\langle         -2\alpha^{8/3}        {\widehat        {\frac{\partial
\widetilde{c}}{\partial  x_i}}}^2  \rangle$,  and  $a_{10}  =  \langle
\alpha^{16/3}    {\widehat   {\frac{\partial   \widetilde{c}}{\partial
x_i}}}^2 \rangle$.  Here, $K_i =
\left(\overline{\widetilde{u_i}\widetilde{c}}                         -
\overline{\widetilde{u_i}}~\overline{\widetilde{c}} \right)$, and $K_i'
= \left(\widehat{\widetilde{u_i}\widetilde{c}} -
\widehat{\widetilde{u_i}}~\widehat{\widetilde{c}} \right)$.


\begin{thebibliography}{99}

\bibitem {germ91}
Germano,~M., Piomelli,~U., Moin,~P.  and Cabot,~W.~H.: 1991, \newblock
{A  dynamic  subgrid-scale   eddy  viscosity  model},  \newblock  {\em
Phys. Fluids A} \textbf{3}, 1760--1765.

\bibitem {pope04} Pope,~S.~B.:
2004, \newblock {Ten questions concerning the large-eddy simulation of
turbulent flows}, \newblock {\em New J. Phys.}  \textbf{6}, 1--24.

\bibitem {meye05}
Meyers,~J.,   Geurts,~B.~J.    and   Baelmans,~M.:   2005,   \newblock
{Optimality of the dynamic procedure for large-eddy simulations},
\newblock {\em Phys. Fluids} \textbf{17}, 045108. 

\bibitem {port00}
Port\'{e}-Agel, F., Meneveau, C.  and Parlange, M. B.: 2000, \newblock
{A   scale-dependent   dynamic   model  for   large-eddy   simulation:
application to  a neutral atmospheric boundary  layer}, \newblock {\em
J. Fluid Mech.} \textbf{415}, 261--284.

\bibitem {port04}
Port\'{e}-Agel, F.:  2004, \newblock {A  scale-dependent dynamic model
for scalar transport in LES of the atmospheric boundary layer},
\newblock {\em Boundary-Layer Meteorol.} \textbf{112}, 81-105.

\bibitem {basu06a}   Basu,~S.   and  Port\'{e}-Agel,~F.:
2006,   \newblock   {Large-eddy   simulation  of   stably   stratified
atmospheric  boundary  layer  turbulence:  a  scale-dependent  dynamic
modeling  approach}, \newblock  {\em J.   Atmos.   Sci.}  \textbf{63},
2074--2091.

\bibitem {basu06b}
Basu,~S.,  Port\'{e}-Agel,~F.,  Foufoula-Georgiou,~E.,  Vinuesa,~J.-F.
and  Pahlow,  M.:  2006,   \newblock  {Revisiting  the  local  scaling
hypothesis in stably stratified atmospheric boundary layer turbulence:
an integration  of field  and laboratory measurements  with large-eddy
simulations},     \newblock     {\em     Boundary-Layer     Meteorol.}
10.1007/s10546-005-9036-2.

\bibitem {bouz05}
Bou-zeid,~E.,  Meneveau,~C.   and  Parlange,~M.:  2006,  \newblock  {A
scale-dependent Lagrangian dynamic model  for large eddy simulation of
complex  turbulent flows}, \newblock  {\em Phys.  Fluids} \textbf{17},
025105.

\bibitem {stol06}   Stoll,~R.   and  Port\'{e}-Agel,~F.:
2006, \newblock {Dynamic subgrid-scale  models for momentum and scalar
fluxes in  large-eddy simulations of  neutrally stratified atmospheric
boundary  layers  over heterogeneous  terrain},  \newblock {\em  Water
Resour. Res.}  \textbf{42}, W01409.

\bibitem {smag63}
Smagorinsky, J.: 1963, \newblock {General Circulation Experiments with
the Primitive Equations}, \newblock {\em Mon. Weath.  Rev.}
\textbf{91}, 99--164.

\bibitem {wong94}
Wong,~V. and  Lilly,~D.: 1994, \newblock {A comparison  of two dynamic
subgrid scale closure methods for turbulent thermal convection},
\newblock {\em Phys. Fluids} \textbf{6}, 1016--1023.

\bibitem {andr94}
Andr\'{e}n,~A.,  Brown,~A.~R.,  Graf,~J., Mason,~P.~J.,  Moeng,~C.-H.,
Nieuwstadt,~F.~T.~M.   and Schumann,~U.:  1994,  \newblock {Large-eddy
simulation of  a neutrally stratified boundary layer:  a comparison of
four codes}, \newblock {\em Q.  J.  Royal Meteorol.  Soc.}
\textbf{120}, 1457--1484.

\bibitem {koso97}
Kosovi\'{c},~B.:  1997,  \newblock  {Subgrid-scale modelling  for  the
large-eddy simulation of high-Reynolds-number boundary layers},
\newblock {\em J. Fluid Mech.} \textbf{336}, 151--182.

\bibitem {chow05}
Chow,~F.~K., Street,~R.~L., Xue,~M. and Ferziger,~J.~H.: 2005,
\newblock {Explicit filtering  and reconstruction turbulence modeling
for large-eddy  simulation of neutral boundary  layer flow}, \newblock
{\em J. Atmos. Sci.} \textbf{62}, 2058--2077.

\bibitem {maso92}
Mason,~P.~J.    and   Thomson,~D.~J.:   1992,  \newblock   {Stochastic
backscatter in  large-eddy simulations of  boundary layers}, \newblock
{\em J. Fluid Mech.} \textbf{242}, 51--78.

\bibitem {sull94}
Sullivan,~P.~P., McWilliams,~J.~C.   and Moeng,~C.-H.: 1994, \newblock
{A  subgrid-scale   model  for  large-eddy   simulation  of  planetary
boundary-layer flows}, \newblock {\em Boundary-Layer Meteorol.}
\textbf{71}, 247--276.

\bibitem {esau04}  Esau, I.: 2004,
\newblock {Simulation  of Ekman boundary  layers by large  eddy model
with dynamic  mixed subfilter closure}, \newblock  {\em Environ. Fluid
Mech.} \textbf{4}, 273--303.

\bibitem {busi71}
Businger,~J.~A., Wyngaard,~J.~C., Izumi,~Y. and Bradley,~E.~F.: 1971,
\newblock  {Flux-profile  relationships  in the  atmospheric  surface
layer}, \newblock {\em J.  Atmos.  Sci.}  \textbf{28}, 181--189.

\bibitem {kade90}
Kader,~B.~A.   and  Yaglom,~A.~M.:  1990,  \newblock {Mean  field  and
fluctuation moments in unstably stratified turbulent boundary layers},
\newblock {\em J. Fluid Mech.} \textbf{212}, 637--662.

\bibitem {gran92}
Grant,~A.~L.~M.: 1992,  \newblock {The structure of  turbulence in the
near-neutral    atmospheric    boundary    layer},   \newblock    {\em
J. Atmos. Sci.} \textbf{49}, 226--239.

\bibitem {gran86}
Grant,~A.~L.~M.:  1986,  \newblock  {Observations  of  boundary  layer
structure   made  during  the   KONTUR  experiment},   \newblock  {\em
Q. J. Royal Meteorol. Soc.} \textbf{112}, 825--841.

\bibitem {warh00}   Warhaft,~Z.:
2000, \newblock  {Passive scalars in turbulent  flows}, \newblock {\em
Annu. Rev. Fluid Mech.} \textbf{32}, 203--240.

\bibitem {shra00}
Shraiman,~B.~I.    and    Siggia,~E.~D.:   2000,   \newblock   {Scalar
turbulence}, \newblock {\em Nature} \textbf{405}, 639--646.

\bibitem {maso89}       Mason,~P.:
1989, \newblock  {Large-eddy simulation of  the convective atmospheric
boundary  layer},  \newblock  {\em  J.   Atmos.   Sci.}   \textbf{46},
1492--1516.

\bibitem {maso87}
Mason,~P.~J.    and   Thomson,~D.~J.:   1987,  \newblock   {Large-eddy
simulations of the neutral-static-stability planetary boundary layer},
\newblock  {\em   Quart.   J.  Roy.    Meteorol.  Soc.}  \textbf{113},
413--443.

\bibitem {moen94}
Moeng,~C.-H.  and Sullivan,~P.~P.:  1994, \newblock  {A  comparison of
shear- and buoyancy-driven  planetary boundary layer flows}, \newblock
{\em J. Atmos. Sci.} \textbf{51}, 999--1022.

\bibitem {ding01} Ding,~F.,
Arya,~S.~P. and Lin,~Y.-L.:  2001, \newblock {Large-eddy simulations
of  the   atmospheric  boundary   layer  using  a   new  subgrid-scale
model. Part  I: Slightly unstable and neutral  cases}, \newblock {\em
Environ. Fluid Mech.} \textbf{1}, 29--47.

\bibitem {carl02}     Carlotti,
P.: 2002,  \newblock {Two-point properties  of atmospheric turbulence
very close  to the  ground: Comparison of  a high resolution  LES with
theoretical models}, \newblock {\em Boundary-Layer Meteorol.}
\textbf{104}, 381--410.

\bibitem {sull03}
Sullivan,~P.~P.,  Horst,~T.~W.,   Lenschow,~D.~H.,  Moeng,~C.-H.  and
Weil,~J.~C.: 2003, \newblock  {Structure of subfilter-scale fluxes in
the   atmospheric  surface  layer   with  application   to  large-eddy
simulation modelling}, \newblock  {\em J.  Fluid Mech.} \textbf{482},
101--139.

\bibitem {lill92} Lilly,~D.~K.: 1992,  \newblock {A proposed modification 
of the Germano subgridscale closure method}, \newblock {\em Phys. Fluids A}
\textbf{4}, 633--635.

\end{thebibliography}
\end{document}